\documentclass[useAMS,usenatbib]{mn2e}
\usepackage{epsfig}
\usepackage{graphics}
\usepackage[usenames]{color}
\usepackage{endnotes}

\usepackage{amssymb}
\usepackage{amsmath}
\usepackage{times}
\usepackage{subfigure}
\newcommand{\be}{\begin{equation}}
\newcommand{\ee}{\end{equation}}
\newcommand{\bdm}{\begin{displaymath}}
\newcommand{\edm}{\end{displaymath}}
\newcommand{\bea}{\begin{eqnarray}}
\newcommand{\eea}{\end{eqnarray}}

\newcommand{\alb}{\textcolor{black}}

\newcommand{\msun}{M_\odot}
\def\lsim{\lower.5ex\hbox{$\; \buildrel < \over \sim \;$}}

\title[MBHB mass bias and pulsar timing arrays]
{\textbf{Selection bias in dynamically-measured super-massive black hole samples:} consequences for pulsar timing arrays}

\author[A. Sesana et al.]
       {Alberto Sesana$^{1}$\thanks{E-mail: asesana@star.sr.bham.ac.uk}, Francesco Shankar$^{2}$, Mariangela Bernardi$^{3}$ \& Ravi K. Sheth$^{3}$ 
\\
$^{1}$ School of Physics and Astronomy, University of
Birmingham, Edgbaston, Birmingham B15 2TT, United Kingdom\\
$^{2}$ Department of Physics and Astronomy, University of Southampton, Highfield, SO17 1BJ, UK\\
$^{3}$ Department of Physics and Astronomy, University of Pennsylvania, 209 South 33rd St, Philadelphia, PA 19104\\
}

\begin{document}

\date{}

\pagerange{\pageref{firstpage}--\pageref{lastpage}} \pubyear{2016}

\maketitle

\label{firstpage}

\begin{abstract}
 Supermassive black hole -- host galaxy relations are key to the computation of the expected gravitational wave background (GWB) in the pulsar timing array (PTA) frequency band. It has been recently pointed out that standard relations adopted in GWB computations are in fact biased-high. We show that when this selection bias is taken into account, the expected GWB in the PTA band is a factor of about three smaller than previously estimated. Compared to other scaling relations recently published in the literature, the median amplitude of the signal at $f=1$yr$^{-1}$ drops from $1.3\times10^{-15}$ to $4\times10^{-16}$. Although this solves any potential tension between theoretical predictions and recent PTA limits without invoking other dynamical effects (such as stalling, eccentricity or strong coupling with the galactic environment), \alb{it also makes the GWB detection more challenging.} 
 

\end{abstract}


\section{Introduction}
The first gravitational wave (GW) detection by advanced LIGO \cite{2016PhRvL.116f1102A} put the field of GW astronomy in the spotlight. At nHz frequency (inaccessible to ground based interferometers), pulsar timing arrays (PTAs) brings the promise of GW detection from inspiralling supermassive black hole (SMBH) binaries. This is the primary goal of the European Pulsar Timing Array \cite[EPTA,~][]{2016arXiv160208511D}, the Parkes Pulsar Timing Array \cite[PPTA,~][]{2013PASA...30...17M} and the North American Nanohertz Observatory for Gravitational Waves \cite[NANOGrav,~][]{2015ApJ...813...65T}, that join forces under the aegis of the International Pulsar Timing Array \cite[IPTA,~][]{2016MNRAS.tmp..131V}.

At those low frequencies, the superposition of individual signal coming from SMBH binaries is expected to produce a stochastic gravitational wave background \citep[GWB][]{sesana08}. Its normalization is set by the cosmic merger rates of SMBH binary systems and their typical masses, whereas its shape is affected by the interaction of the binaries with their stellar and gaseous environment, possibly suppressing the signal at the lowest frequencies. Both the amplitude and the spectral shape of the signal are affected by significant uncertainties, some of which have been recently explored by several authors \citep[see, e.g.][]{KS11,sesana13,sesana13CQGra,ravi14}. In particular the signal amplitude strongly depends on the SMBH masses, which is set by the intrinsic relation between SMBHs and either the host galaxy stellar bulge ($M_{\bullet}-M_b$ relation) or the stellar velocity dispersion ($M_{\bullet}-\sigma$ relation). In recent years, improvements in SMBH dynamical mass measurements, together with the discovery of few ``overmassive'' black holes in brightest cluster galaxies \citep{2013ApJ...764..184M} resulted in a upward revision of the SMBH-host galaxy relations \citep[][hereinafter KH13]{2013ARA&A..51..511K}, implying a median expected GWB signal with strain amplitude $A\sim10^{-15}$ at $f=1/$yr (although values of $A<10^{-15}$ are not excluded). Therefore, recent PTA non detections of the GWB at the $A\sim10^{-15}$ level \citep{2015Sci...349.1522S} has been interpreted as being somewhat in tension with currently favoured SMBH assembly scenarios, pointing to a possible important role of SMBH binary eccentricity or environmental coupling \citep{2015arXiv150803024A}.

However, recently \alb{\cite{2016arXiv160301276S} (S16, hereafter), confirming the earlier finding of \cite{2007ApJ...660..267B}, showed that the set of local galaxies with dynamically-measured SMBHs is biased. It has significantly higher velocity dispersions than local galaxies of similar stellar mass as determined from the Sloan Digital Sky Survey \citep[e.g.][]{2014MNRAS.443..874B} and this bias also affects the SMBH - host galaxy relations. Using targeted Monte-Carlo simulations, S16 showed that this bias could have been induced by the observational selection requirement that the black hole sphere of influence must be resolved to measure black hole masses with spatially resolved kinematics. By studying the impact of this bias on the local SMBH scaling relations, they found this selection effect to artificially increase the normalization of the intrinsic $M_{\bullet}-\sigma$ relation by a factor $\gtrsim 3$, and the intrinsic $M_{\bullet}-M_b$ relation by up to an order of magnitude at low stellar masses. The underlying unbiased relations would thus lie significantly below all previous estimates found in the literature, naturally implying a lower normalization of the expected GWB.}

The aim of this Letter is to quantify the impact of the bias in local SMBH scaling relations on the detectability of the GWB with PTAs. In Section 2 we review our model for the computation of the GWB, and in Section 3 we focus on the SMBH-host galaxy scaling relations. We present our results in Section 4, discussing their implication for PTA campaigns and we summarize our findings in Section 5. Throughout the paper we assume a concordance $\Lambda$--CDM universe with $h=0.7$, $\Omega_M=0.3$ and $\Omega_\lambda=0.7$. Unless otherwise specified, we use geometric units where $G=c=1$.

\section{Gravitational wave background model}
The method we adopt to extract the GWB from observed properties of low redshift galaxies is fully described in \cite{sesana13} (S13, hereafter). In the following we provide a short summary. We neglect issues related to possible binary coupling with the environment, eccentricity, and the possibility of resolving individual sources, which are beyond the scope of this Letter. We consider a cosmological population of SMBH binaries, inspiralling in quasi-circular orbit, driven by GW back-reaction. Each merging pair is characterized by the masses of the two black holes $M_{\bullet,1}>M_{\bullet,2}$, defining the mass ratio $q_\bullet=M_{\bullet,2}/M_{\bullet,1}${\footnote{We use $M_{\bullet,1},M_{\bullet,2},q_\bullet$ for SMBH binaries, and $M$ and $q$ for galaxies}}. The characteristic amplitude $h_c$ of the generated GWB is given by \cite{sesana08}
\begin{equation}
h_c^2(f) =\frac{4}{\pi f^2}\int \int \int
dzdM_{\bullet,1}dq_\bullet \, \frac{d^3n}{dzdM_{\bullet,1}dq_\bullet,}
{1\over{1+z}}~{{dE_{\rm gw}({\cal M})} \over {d\ln{f_r}}}\,.
\label{hcdE}
\end{equation}
Although the integrals formally run through the whole allowed range of each variable, \cite{sesana08,sesana13} showed that $>95\%$ of the signal comes from major mergers ($q_\bullet>1/4$) involving massive binaries ($M_{\bullet,1}>10^8\msun$) at low redshift ($z<1.5$). The energy emitted per log-frequency interval is 
\begin{equation}
\frac{dE_{\rm gw}}{d\ln{f_r}}=\frac{\pi^{2/3}}{3}{\cal M}^{5/3}f_r^{2/3}\,.
\label{dedlnf}
\end{equation}
Here ${\cal M}=(M_{\bullet,1}M_{\bullet,2})^{3/5}/(M_{\bullet,1}+M_{\bullet,2})^{1/5}$ is the chirp mass of the binary and $f_r=(1+z)f$ is the GW rest frame frequency (twice the orbital frequency). It then follows \citep{jen06} that
\begin{equation}
h_c(f) =A\, \left(\frac{f}{{\rm yr}^{-1}}\right)^{-2/3}\,
\label{hcpar}
\end{equation}
where the normalization constant $A$ depends on the SMBH binary merger rate density per unit redshift, mass and mass ratio given by the term $d^3n/(dzdM_{\bullet,1}dq_\bullet)$ in equation (\ref{hcdE}). PTA limits on a stochastic GWB are usually quoted in terms of $A$ \citep{2015MNRAS.453.2576L,2015arXiv150803024A,2015Sci...349.1522S}.

\subsection{Massive black hole binary merger rate}
We summarize here the approach taken by S13 to determine $d^3n/dzdM_{\bullet,1}dq_\bullet$. The procedure is twofold: (i) we determine {\it from observations} the {\it galaxy} merger rate $d^3n_G/dzdMdq$ (in a merging galaxy pair, $M$ and $q<1$ are the stellar mass of the primary galaxy and the mass ratio respectively), and (ii) we populate merging galaxies with SMBHs according to empirical SMBH -- host galaxy scaling relations. \alb{In this procedure, we assume a one-to-one correspondence between galaxy and SMBH mergers; significant delays between the two or possible stalling of SMBH pairs will naturally {\it reduce} the integrated GWB.}

\subsubsection{Galaxy merger rate}
The galaxy differential merger rate can be written as (S13)
\begin{equation}
\frac{d^3n_G}{dzdMdq}=\frac{\phi(M,z)}{M\ln{10}}\frac{{\cal F}(z,M,q)}{\tau(z,M,q)}\frac{dt_r}{dz}.
\label{galmrate}
\end{equation}
$\phi(M,z)=(dn/d{\rm log}M)_z$ denotes the galaxy mass function (MF) at redshift $z$; ${\cal F}(M,q,z)=(df/dq)_{M,z}$ is the differential fraction of galaxies with mass $M$ at redshift $z$ paired to a secondary galaxy with mass ratio in the range $q, q+\delta{q}$; $\tau(z,M,q)$ is the merger time-scale for a galaxy pair and is a function of $M$, $q$ and $z$. $dt_r/dz$ provides the conversion between proper time and redshift in the adopted cosmology. Equation (\ref{galmrate}) conveniently expresses the rate as a function of the directly observable quantities $\phi$ and ${\cal F}$, whereas $\tau$ can be inferred from numerical simulations (see below).

\begin{figure}
\includegraphics[scale=0.43,clip=true,angle=0]{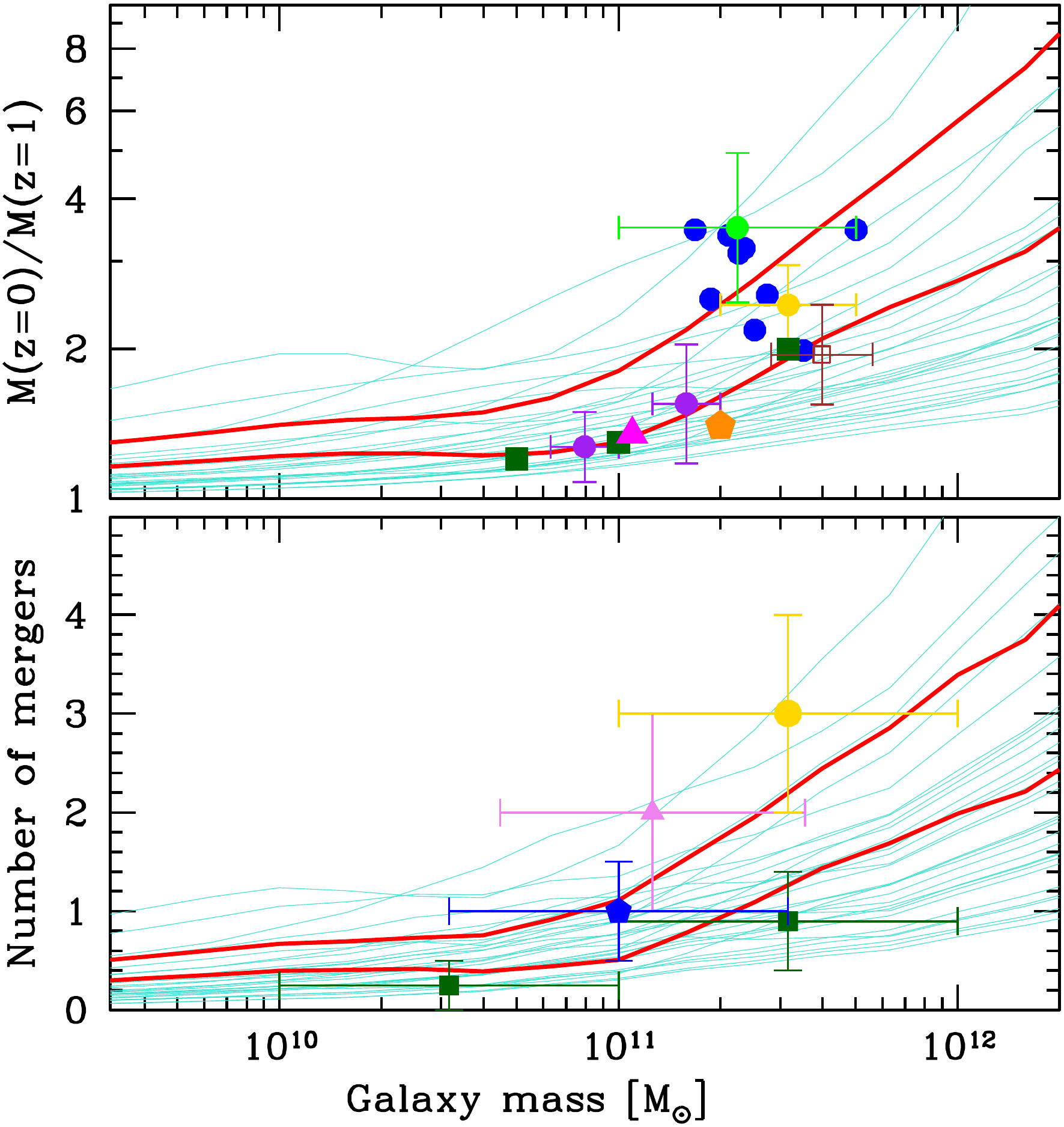}
\caption{Top panel: average galaxy stellar mass growth, $M(z=0)/M(z=1)$, predicted by all combinations of pair fractions and merger time-scales considered in this work (light blue lines), compared to a compilation of theoretical and observational estimates. Dark green squares are simulations from \protect\cite{oser10}, the magenta triangle is from \protect\cite{naab09}, the blue circles are a compilation of BCG simulations by \protect\cite{laporte13} \alb{and the brown open square is from \protect\cite{2015ApJ...802...73S}}. Observational estimates are from \protect\cite{mcintosh08} (purple circles), \protect\cite{vandokkum10} (orange pentagon) and \protect\cite{lidman12} (yellow circle). Red lines highlight models featuring an average mass growth in line with what observed for BCGs (upper), and with typical massive galaxies (lower). Bottom panel, average number of mergers as a function of galaxy mass since $z=1$. Light blue lines are the models presented in this paper, with fiducial ones highlighted in red. Dark green square are estimates from \protect\cite{hopkins10}, the blue pentagon from \protect\cite{xu12}, the violet triangle from \protect\cite{delucia07} and the yellow circle from \protect\cite{trujillo11}.}


\label{fig1}
\end{figure}

Improving on S13, we explore four different mass functions (labelled, MF1-4). MF1-3 are constructed by matching individually the MFs at $z>0$ provided by \cite{2013A&A...556A..55I,muzzin13,tomczak14} to the local mass function estimated by \cite{2013MNRAS.436..697B}. Conversely MF4 simply assumes an extension of \cite{2013MNRAS.436..697B} at all considered redshifts without evolution (therefore providing an upper limit to the inferred GWB). This is also in line with the recent observational findings by \cite{2016MNRAS.455.4122B} at $z\sim 0.5$. We also explore four different differential pair fractions (PFs) $=(df/dq)_{M,z}$, extracted from \citep{bundy09,deravel09,lopez12,xu12} following the same procedure as in S13 \citep[see also detailed description in][]{2015MNRAS.446...38G}. We checked that those broadly match recent compilations of galaxy PFs from, e.g, \cite{2014MNRAS.444.1125C}. Considering different MFs and PFs helps in folding into our computation possible systematic errors due to the specific samples and techniques adopted in each work. To account for statistical uncertainties, each MF and PF comes with a fiducial value plus an upper and lower limits derived from the uncertainty range quoted in each paper. As in S13, we explore a 'fast' and 'slow' merger scenario where $\tau$ is specified by (i) adopting equation (10) of \cite{kit08}
\begin{equation}
\tau = 1.32\,{\rm Gyr}\left( \frac{M}{4\times10^{10} h^{-1} M_\odot} \right)^{-0.3}\left(1+\frac{z}{8}\right),
\label{tau}
\end{equation}
and (ii) complementing equation (\ref{tau}) with fits to the results of a set of full hydrodynamical simulation of galaxy mergers presented by \cite{lotz10}, which gives
\begin{equation}
\tau = 0.79\,{\rm Gyr}\left( \frac{M}{4\times10^{10} h^{-1} M_\odot} \right)^{0.3}q^{-0.1}\left(1+\frac{z}{8}\right).
\label{taulotz}
\end{equation}

We interpolate all the measured $\phi,{\cal F},\tau$ on a fine 3-D grid in $(z,M,q)$, to numerically obtain $12\times12\times2=288$ differential galaxy merger rates. Note that typical values of $\tau$ are up to few Gyr, therefore, the merger rate at a given $(z,M,q)$ point in the grid is obtained by evaluating $\phi$ and ${\cal F}$ at $(z+\delta{z},M,q)$, where $\delta{z}$ is the redshift delay corresponding to the merging time $\tau$. By doing this, we implicitly assume that SMBH binaries coalesce instantaneously at the merger time of their hosts. We extend our calculations to all galaxies with $M>10^{10}\msun$ merging at $z<1.3$, ensuring that we capture the bulk of the GWB \citep[see][]{sesana13}.

Using the same formalism of equation \ref{galmrate} and following \cite{2015MNRAS.446...38G}, the differential number of mergers experienced by each individual galaxy with mass $M$ is given by
\begin{equation}
\frac{d^2N}{dzdq}\bigg\rvert_M=\frac{df}{dq}\bigg\rvert_{M,z}\frac{1}{\tau(z,M,q)}\frac{dt_r}{dz}.
\label{galmrateind}
\end{equation}
We can then compute the number of mergers experienced by each individual galaxy with a given mass at $z=1$ ($M_{z=1}$) as
\begin{equation}
N({M_{z=1}})=\int_1^0dz\int_{q_{\rm min}}^1 dq\int dM \frac{d^2N}{dzdq}\bigg\rvert_M\delta[M-M(z)],
\label{Nmerger}
\end{equation}
where the integral is consistently evaluated at the redshift-evolving galaxy mass $M(z)$ through the Dirac delta function. If we multiply the integrand of equation (\ref{Nmerger}) by a factor $qM$, we then obtain the mass growth of the galaxy, $M_{z=0}/M_{z=1}$, since $z=1$. Results for all MFs and PFs considered in this work are compared to various estimates from the literature in figure \ref{fig1}.
The average number of galaxy mergers and mass growth are in line with observations and other theoretical estimates, validating the viability of our modelling.


\section{SMBH-galaxy scaling relations}

The last ingredient in the computation of the GWB from equation (\ref{hcdE}) is the connection between the host galaxy and the SMBH mass. This issue is overcome by employing scaling relations between $M_\bullet$ and either the galaxy bulge stellar mass $M_b$ or velocity dispersion $\sigma$, measured on a limited sample of nearby galaxies. Recent estimates of the GWB are usually based on relations provided by \cite{mcconnell12} and KH13. In particular, we consider here the relations reported by KH13:
\begin{equation}
{\rm log}_{10} \left(\frac{M_\bullet}{\msun}\right)=8.5+4.42{\rm log}_{10}\left(\frac{\sigma}{200{\rm km/s}}\right),
\label{srelkh13sigma}
\end{equation}
\begin{equation}
{\rm log}_{10} \left(\frac{M_\bullet}{\msun}\right)=8.69+1.17{\rm log}_{10}\left(\frac{M_b}{10^{11}\msun}\right),
\label{srelkh13bulge}
\end{equation}
with intrinsic scatter $\epsilon=0.29$ dex, and $\epsilon=0.3$ dex respectively.

However, S16 have shown that both Monte Carlo simulations and the analysis of the residuals around the scaling relations suggest that $\sigma$ is most fundamental galaxy property correlated to SMBH mass, with a possible additional (weak) dependence on stellar mass. In particular the $M_{\bullet}-M_b$ relation is mostly induced by the relation between $M_{\bullet}$-$\sigma$ and $\sigma-M_b$. S16 find the preferred intrinsic relation
\begin{equation}
{\rm log}_{10} \left(\frac{M_\bullet}{\msun}\right)=7.7+5{\rm log}_{10}\left(\frac{\sigma}{200{\rm km/s}}\right)+0.5{\rm log}_{10}\left(\frac{M_b}{10^{11}\msun}\right),
\label{srels16}
\end{equation}
with intrinsic scatter $\epsilon=0.25$ dex. In the following we will consider the three scaling relations given above; models adopting either equation (\ref{srelkh13sigma}) or (\ref{srelkh13bulge}) will be referred to as KH13, whereas those adopting equation (\ref{srels16}) will be referred to as S16.

The relations link $M_{\bullet}$ to the {\it bulge} properties, whereas our galaxy merger rates are function of the total stellar mass. We derive the bulge mass of each galaxy by multiplying the total stellar mass by a factor $f_b$ taken from \cite{2014MNRAS.443..874B} for SerExp galaxies with a probability $P>0.5$ of being ellipticals/lenticulars, which we fit as
\begin{eqnarray}
 f_b &=&
\begin{cases}
     0.018({\rm log}_{10}M_b-9.5) & 9.5 \le {\rm log}_{10}M_b < 10.6 \\
     0.175({\rm log}_{10}M_b-10.6) & 10.6 \le {\rm log}_{10}M_b < 11.4 \\
     0.004({\rm log}_{10}M_b-11.4) & {\rm log}_{10}M_b \ge 11.4 \\
\end{cases}
\label{fbulge}
\end{eqnarray}
with an intrinsic scatter $\epsilon=0.2$ dex.
Velocity dispersion $\sigma$, corrected to the Hyperleda aperture of $0.595$ kpc for consistency with S16, is computed from the corresponding bulge mass and fitted as
\begin{equation}
{\rm log}_{10} \sigma =2.58-0.087({\rm log}_{10}M_b-12.9)^2, 
\end{equation}
characterized by an intrinsic scatter
\begin{equation}
  \epsilon =0.044-0.015({\rm log}_{10}M_b-12.5)^2 {\rm dex}.
  \label{sigmadisp}
\end{equation}
In equations (\ref{fbulge})-(\ref{sigmadisp}), $M_b$ and $\sigma$ are expressed in units of $\msun$ and km s$^{-1}$, respectively. We assume no redshift evolution in any of the aforementioned relations up to $z=1.3$.



Finally, each merger event, involves three bulges: the progenitors $M_{\rm b,1}$, $M_{\rm b,2}$ and the remnant $M_{\rm b,r}$. We associate to these spheroids SMBH masses $M_{\bullet,1}$, $M_{\bullet,2}$ and $M_{\bullet,r}$, taken from the same scaling relation, and imposing the only constrain that $M_{\bullet,1}+M_{\bullet,2}<M_{\bullet,r}$. We therefore allow, during the merger, an amount of accretion $M_{\rm acc}=M_{\bullet,r}-(M_{\bullet,1}+M_{\bullet,2}<M_{\bullet,r})$. This mass can be accreted with a different timing with respect to the SMBH binary merger, and in different amount on the two SMBHs. We follow the three prescriptions described in Section 2.2 of \cite{sesana09}. 
We stress that the exact details of the accretion model are not crucial as we are specifically interested in evaluating the impact of the bias in SMBH scaling relations for PTA searches.

The combination of the $3\times3=9$ SMBH population prescription with the 288 galaxy merger rates results in 2592 different SMBH binary merger rates $d^3n/dzdM_{\bullet,1}dq_\bullet$. By construction they are consistent with current observations of the evolution of the galaxy mass function, pair fractions at $z<1.3$ and $M>10^{10}\msun$, and with published empirical SMBH-host relations. They also reproduce {\it observational constraints} on the number of galaxy mergers and mass growth (cf. figure \ref{fig1}).

\section{Results: impact on the expected GWB}
\begin{figure}
\includegraphics[scale=0.4,clip=true,angle=0]{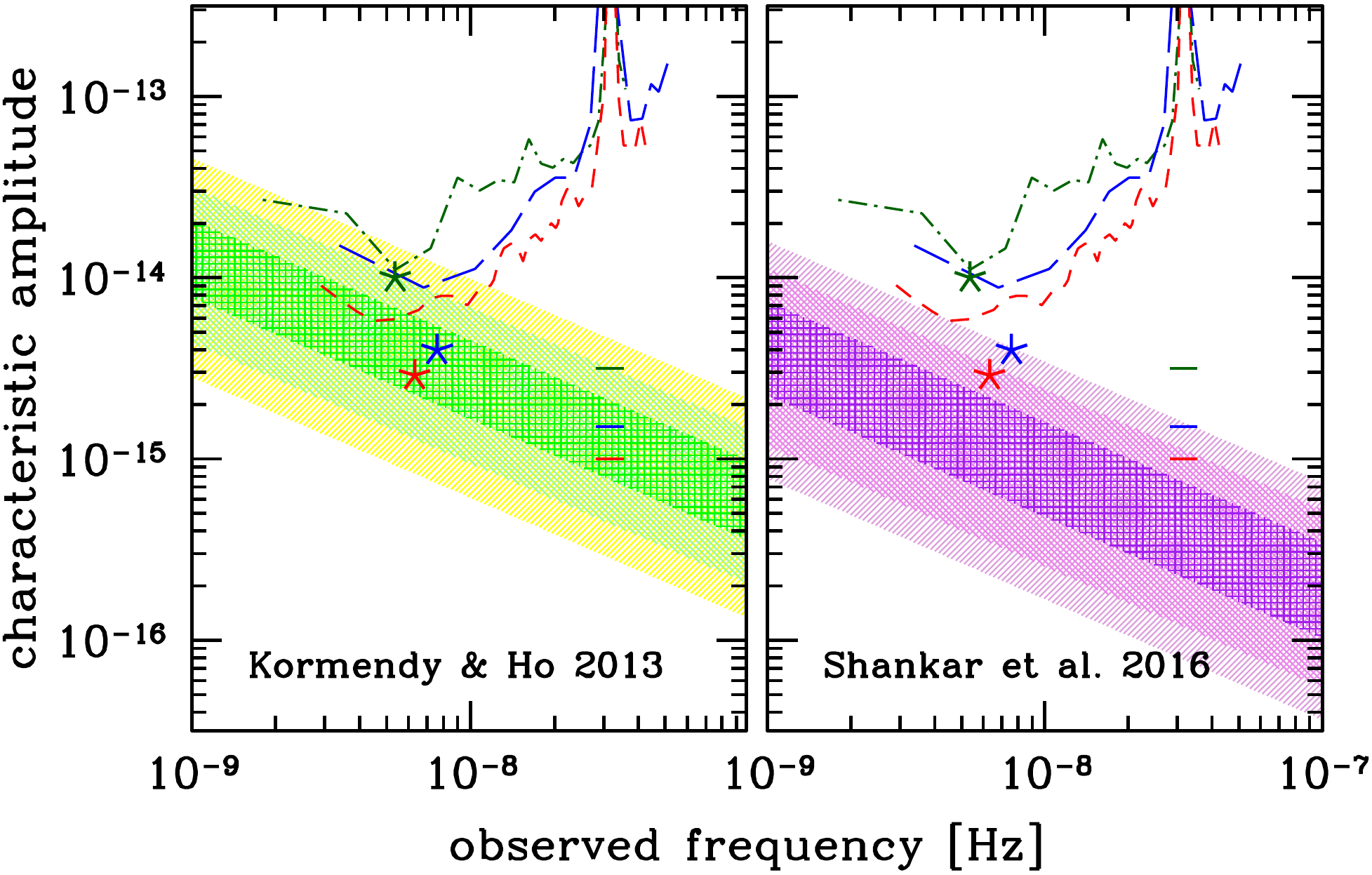}
\caption{Characteristic amplitude of the GWB assuming the scaling relations indicated in the panels. In each panel, the shaded areas represent the 68\% 95\% and 99.7\% 
 confidence intervals of the signal amplitude. The jagged curves are current PTA sensitivities: EPTA (dot-dashed green), NANOGrav (long-dashed blue), and PPTA (short-dashed red). For each sensitivity curve, stars represent the integrated upper limits to an $f^{-2/3}$ background -- cf. equation (\ref{hcpar}) --, and the horizontal ticks are their extrapolation at $f=1$yr$^{-1}$}
\label{fig2}
\end{figure}

\begin{figure}
\includegraphics[scale=0.48,clip=true,angle=0]{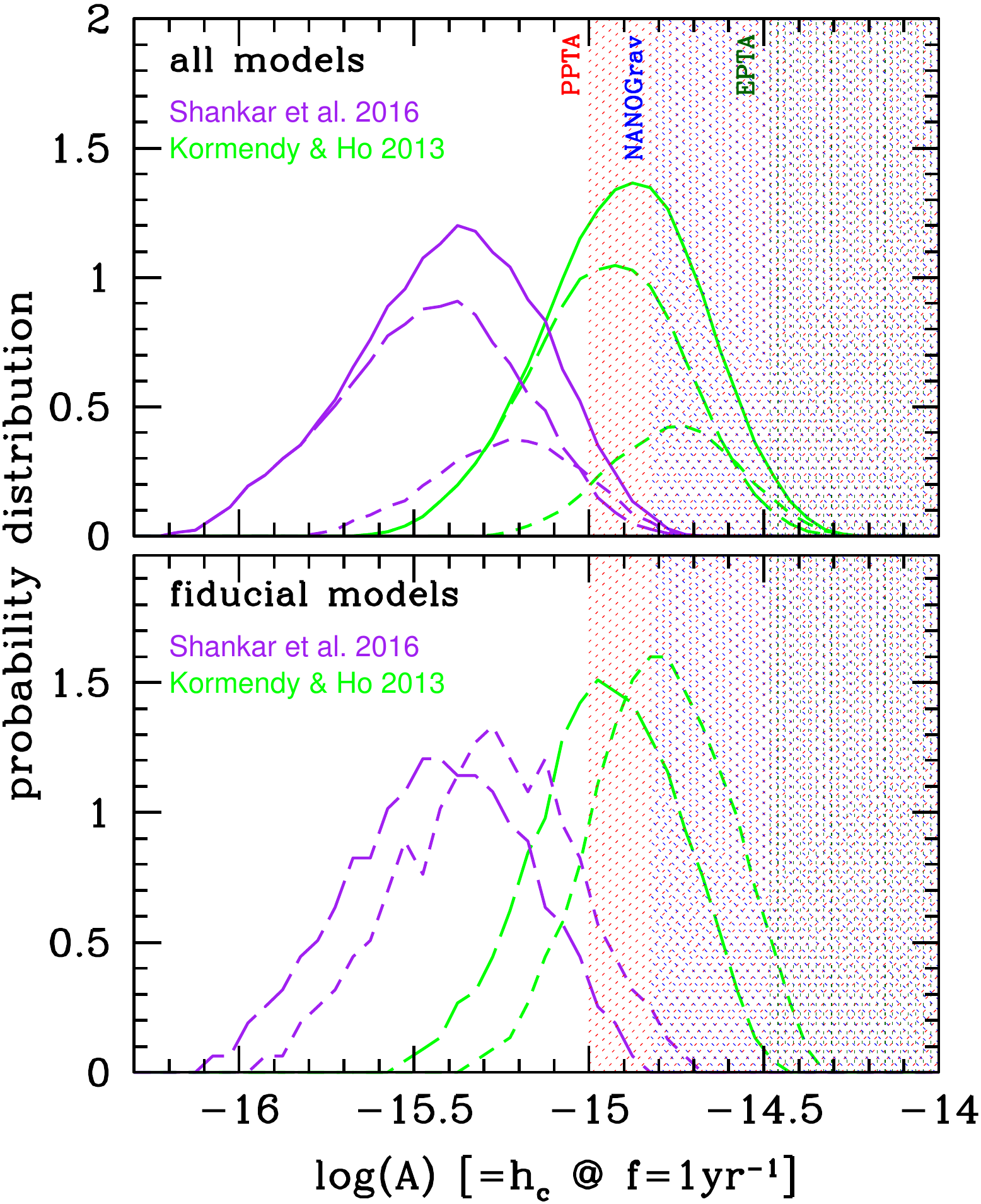}
\caption{Probability distribution of the signal amplitude $A$ assuming different scaling relations, as indicated in figure. Top panel: all pair fractions and merger time-scale are considered. The solid distributions include all the adopted MFs, the long-dashed ones include MF 1-3, and the short-dashed one assumes MF 4. Bottom panel: amplitude distributions assuming the two fiducial models highlighted in figure \ref{fig1}: fiducial1 (long--dashed) and fiducial2 (short--dashed).}
\label{fig3}
\end{figure}

Since we are primarily interested in assessing the impact of the SMBH-galaxy relations selection bias on the expected GWB, we divide the 2592 models in two sub-sample featuring either the KH13 or the S16 relations. The expected characteristic amplitude ranges obtained in the two cases are compared in figure \ref{fig2}, together with current upper limits placed by the three pulsar timing arrays: EPTA \citep{2015MNRAS.453.2576L}, NANOGrav \citep{2015arXiv150803024A} and PPTA \citep{2015Sci...349.1522S}. Including the S16 scaling relations in the model lowers the median value of the expected signal at $f=1$yr$^{-1}$ by a factor of three, from $1.3\times10^{-15}$ to $4\times10^{-16}$. This is further elucidated in figure \ref{fig3}, where we show the resulting probability density distributions of $A$ -- cf. equation (\ref{hcpar}) -- for the two cases. Although recent limits show some tension with the KH13 models, they are fully consistent with the S16 models, which predict $1.4\times10^{-16}<A<1.1\times10^{-15}$ at 95\% confidence. In the upper panel of figure \ref{fig3} we further separate models featuring MF1-3
to those featuring MF4.
As expected, the latter provides an upper limit to the signal, placing the median value at $A=2\times10^{-15}$ and $A=6\times10^{-16}$ for KH13 and S16 models respectively. Conversely, the more realistic MF1-3, place the median signals at $A=1.2\times10^{-15}$ and $A=3.5\times10^{-16}$.


In figure \ref{fig1}, we highlighted in red two 'fiducial' models; one providing a good fit to the overall mass growth of medium-size galaxies (fiducial1, lower red curve), and one matching the mass growth of brightest cluster galaxies (fiducial2, upper curve). The resulting $A$ distributions from these models are reported in the lower panel of figure \ref{fig3}. Obviously, model fiducial2 results in a higher signal, but only by just about 0.2 dex. Median values remain in the range $A=1-2\times10^{-15}$ and $A=3-5\times10^{-16}$ for the KH13 and S16 models respectively. 

\subsection{Implication for PTA detection}
\begin{figure}
\includegraphics[scale=0.4,clip=true,angle=0]{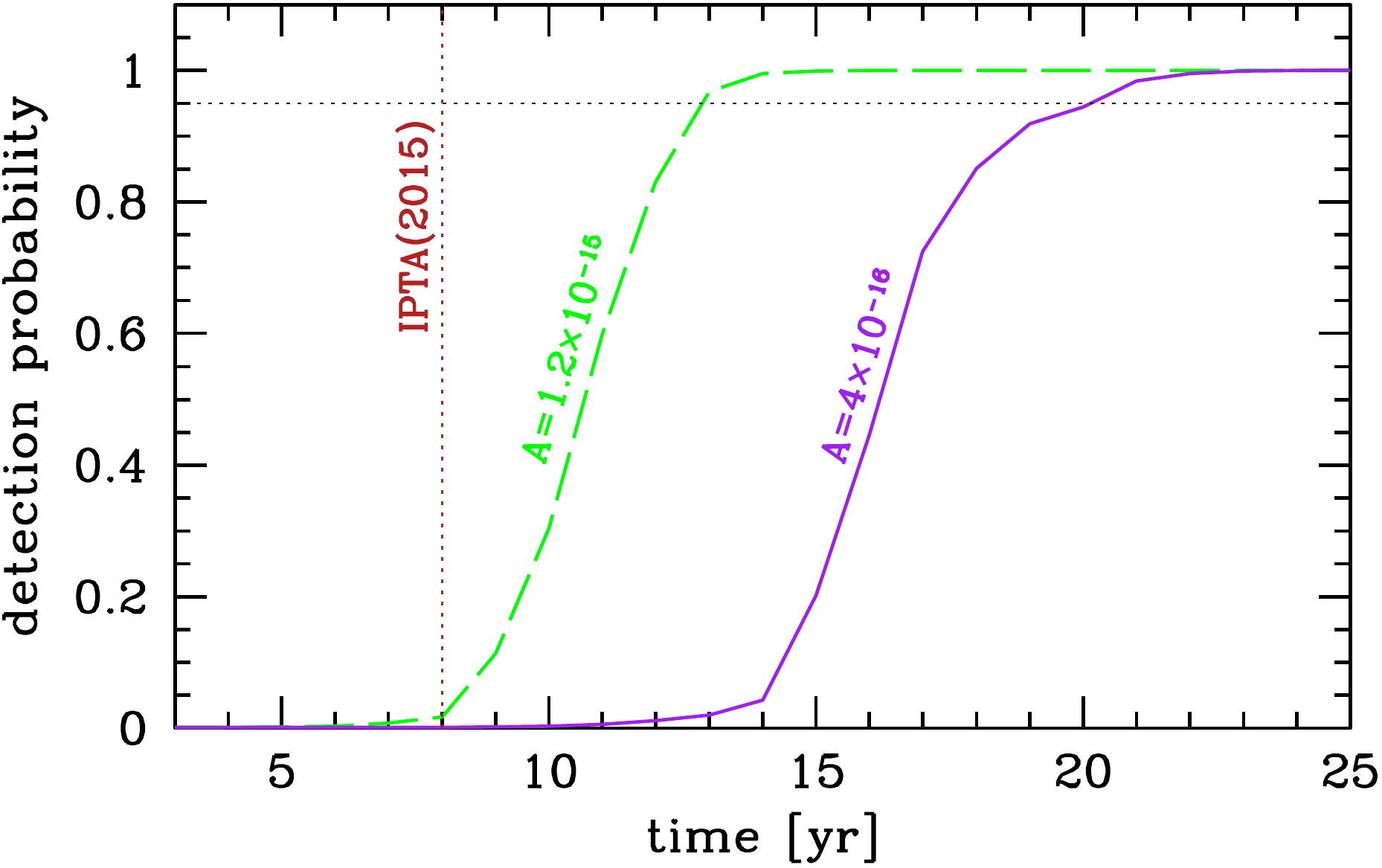}
\caption{Detection probability as a function of time $T$. The dashed green and solid purple lines are for signals with $A=1.2\times10^{-15}$ (KH13) and $A=4\times10^{-16}$ (S16) respectively, assuming an array of $N=50$ pulsars, $\sigma_{\rm rms}=200$ ns and $\Delta{t}=2$ weeks. The black dotted line marks a detection probability of 95\%.}
\label{fig4}
\end{figure}

To investigate the consequences for PTA detection we follow \cite{2015MNRAS.451.2417R} (hereinafter R15). We consider an ideal IPTA-type array with $N=50$ pulsars, timed with an rms residual $\sigma_{\rm rms}=200$ ns at intervals $\Delta{t}=2$ weeks. We compute detection probabilities versus time by means of equation (15) in R15, fixing a false alarm rate $p=0.001$. We mimic the effect of fitting for the pulsar spin-down by excluding the two lowest frequency bins from the computation. Although this is a crude approximation, it has little impact on our investigation, since we are interested in comparative results.  We take two GWBs with $A=1.2\times10^{-15}$ and $A=4\times10^{-16}$, representative of the KH13 and S16 models respectively. Results are shown in figure \ref{fig4}. In the latter case, there is a significant delay in detection probability build-up, reaching 95\% about seven year later. Since, for any given $T$, S/N$\propto N h_c^2/(\sigma_{\rm rms}^2\Delta{t})$, a drop of a factor of three in $h_c$ can be compensated by an equivalent improvement of the rms residuals (i.e. from $\sigma_{\rm rms}=200$ ns to $\sigma_{\rm rms}=70$ ns), by monitoring pulsars every couple of days, or by increasing the number of pulsars in the array.  Note that results in figure \ref{fig4} are shown {\it from the start} of the PTA experiment; for comparison, an $A=1.2\times10^{-15}$ signal would have a detection probability of only few\% in current IPTA data, which is marked by the vertical dotted line. However, one cannot simply extrapolate IPTA detection times based on this plot, since the S/N build-up pace depends on the number of pulsars, on their timing accuracy, on the addition of new pulsars and availability of novel instrumentation.

\section{Conclusions}
We investigated the impact of the selection bias in dynamical measurement of SMBH masses on the expected GWB at nHz frequencies, accessible to PTAs. We found that the revised SMBH-host galaxy relations imply a drop of a factor of three in the signal amplitude, shifting its median to $A=4\times 10^{-16}$. Note that since $A^2$ is proportional to the number of mergers, compensating this drop would require $\approx10$ more SMBH binary mergers, which is severely inconsistent with observational and theoretical estimates of galaxy merger rate (cf. figure \ref{fig1}). This result resolves any potential tension between recent PTA non-detections and theoretical predictions, without invoking any additional physics related to the dynamics of SMBH binaries, such as stalling, high eccentricity or strong coupling with the surrounding stellar and gaseous environment. On the other hand, it also poses a significant challenge to ongoing PTA efforts. If SMBH-galaxy relations are in fact affected by the strong bias reported by S16, the resulting GWB might be outside the reach of current PTAs for several years to come. This picture is consistent with the 'stalling scenario' of \cite{2016ApJ...819L...6T}, in which a 95\% detection probability is expected only in the next 8-to-12 years depending on the array. This is not surprising since their 'stalling scenario' has a mean $A$ consistent with the implication of the selection bias in the SMBH-host relations described here. Our results call for a more extensive investigation along the lines of \cite{2015aska.confE..37J} and \cite{2016ApJ...819L...6T}, properly weighting-in new ultra-precise timing measurement with MeerKAT \citep{2012AfrSk..16..101B} and FAST \citep{2011IJMPD..20..989N} starting next year, and eventually data collected with SKA from 2021 \citep{2009A&A...493.1161S} 

\section*{Acknowledgements}
A.S. is supported by a University Research Fellow of the Royal Society, and acknowledge the continuous support of colleagues in the EPTA. A.S. thanks A. Possenti and S. Taylor for useful comments.

\footnotesize{
  \bibliographystyle{mn2e_Daly}
  \bibliography{references}
}


\bsp

\label{lastpage}

\end{document}